\documentclass[conference]{IEEEtran}
\IEEEoverridecommandlockouts
\usepackage{cite}
\usepackage{amsmath,amssymb,amsfonts}
\usepackage{algorithmic}
\usepackage{graphicx}
\usepackage{textcomp}
\usepackage{xcolor}
\def\BibTeX{{\rm B\kern-.05em{\sc i\kern-.025em b}\kern-.08em
    T\kern-.1667em\lower.7ex\hbox{E}\kern-.125emX}}

\usepackage{mathtools}

\usepackage{steinmetz}

\usepackage{blindtext}
\usepackage{caption}
\usepackage{subcaption}

\begin{document}

\title{A Proof of Concept for OTFS Resilience in Doubly-Selective Channels by GPU-Enabled Real-Time SDR}

\author{
\IEEEauthorblockN{Yi~Xien~Yap, 
		Neil~Bhushan, 
		Onur~Dizdar, 
        Ata~Sattarzadeh,
        David~Redgate,
        Venkateswara~Battula,\\
        and~Stephen~Wang
}
%\IEEEauthorblockA{Research \& Technology Group, VIAVI Solutions, London, UK\\
\IEEEauthorblockA{VIAVI Marconi Labs, VIAVI Solutions, London, UK\\
Email: \{name.surname\}@viavisolutions.com}

%\IEEEauthorblockN{Onur Dizdar}
%\IEEEauthorblockA{\textit{Research \& Technology Group} \\
%\textit{VIAVI Solutions}\\
%London, UK \\
%onur.dizdar@viavisolutions.com}
%\and
%\IEEEauthorblockN{Ata Sattarzadeh}
%\IEEEauthorblockA{\textit{Research \& Technology Group} \\
%\textit{VIAVI Solutions}\\
%London, UK \\
%ata.sattarzadeh@viavisolutions.com}
%\and
%\vspace{0.3cm}
%\IEEEauthorblockN{Stephen Wang}
%\IEEEauthorblockA{\textit{Research \& Technology Group} \\
%\textit{VIAVI Solutions}\\
%London, UK \\
%stephen.wang@viavisolutions.com}
%\vspace{-0.7cm}
%\and
%\IEEEauthorblockN{Yi Xien Yap}
%\IEEEauthorblockA{\textit{Research \& Technology Group} \\
%\textit{VIAVI Solutions}\\
%London, UK \\
%yap.yixien@viavisolutions.com}
}

\maketitle

\begin{abstract}
Orthogonal time frequency space (OTFS) is a modulation technique which is robust against the disruptive effects of doubly-selective channels. 
In this paper, we perform an experimental study of OTFS by a real-time software defined radio (SDR) setup.  
Our SDR consists of a Graphical Processing Unit (GPU) for signal processing programmed using Sionna and TensorFlow, and Universal Software Radio Peripheral (USRP) devices for air interface. We implement a low-latency transceiver structure for OTFS and investigate its performance under various Doppler values. By comparing the performance of OTFS with Orthogonal Frequency Division Multiplexing (OFDM), we demonstrate that OTFS is highly robust against the disruptive effects of doubly-selective channels in a real-time experimental setup.

\end{abstract}


\section{Introduction}
Future communication systems should be able to support different service types and provide reliable communication over a diverse set of channels. One of the most challenging scenarios is  the high-mobility scenario where conventional frequency-based transmissions like OFDM, face severe inter-carrier interference (ICI). The Doppler spread of the the time-variant channels will turn the channel into a doubly selective or doubly dispersive channel. OFDM-based transmission schemes are basically designed to mitigate the time selectivity of the channel and when it comes to a doubly selective channels their performance will degrade. In contrast, recently proposed Orthogonal Time Frequency and Space (OTFS) transmission has been shown to be resilient to delay and doppler shifts of the signal \cite{hadani_2017, hadani_2017_2}. In OTFS, data is modulated in the delay-doppler domain which is the same domain that we usually characterize the wireless communication channel. The interesting property of the OTFS is that localized data in delay-doppler domain can stay localized in time and frequency\cite{hadani_2017}. At the first glance it may seen in contrast with the Heisenberg uncertainty principle; but it can be shown that as long as signal generated in delay-doppler domain is quasi-periodic and have maintain certain relations between the defined regions in delay and doppler, it can be both localized in time and frequency. This property is very important and in fact enables OTFS to be theoretically resilient to selectivity in time and frequency. From practical point of view, OTFS can be implemented as a precoder in the conventional OFDM systems. These properties makes OTFS a proper candidate for high-Doppler scenarios like V2X, Non-terrestrial communications and also where there is uncertainty in the CFO like mm-Wave communications \cite{Hanzo_2022, wei_2021}. In \cite{Surabhi}, the authors presents a formal analysis of OTFS's diversity order in doubly-dispersive channels, demonstrating its performance superiority over OFDM, and confirms through simulation results that OTFS can achieve full diversity in the delay-Doppler domain.

Apart from the theoretical studies, experimental studies have also been conducted to verify the performance of OTFS in realistic environments \cite{thaj_2019, abushattal_2022}. 
In \cite{thaj_2019}, the authors use Universal Software Radio Peripheral (USRP) devices to transmit and receive OTFS signals. The Doppler channel emulation and signal processing is handled in LABView, which uses the host PC Central Processing Unit (CPU). The authors in \cite{abushattal_2022} use  signal generator and analyzer devices to transmit and receive an OTFS signal and use a reverberation chamber to emulate the channel effects. Most existing studies primarily focus on the BER performance comparison between OTFS and OFDM. However, there is a notable lack of discussion regarding the actual overhead of these systems. This oversight is significant as the overhead directly impacts the total throughput of the system. 

In this work, we perform experimental evaluation of OTFS and its resilience in doubly-selective channels in a real-time measurement environment. We implement a full transceiver chain using Universal Software Radio Peripheral (USRP) devices as transmitter and receiver, each connected to a Graphical Processing Unit (GPU) for channel emulation and signal processing. We employ a low-complexity channel estimator and a Message-Passing (MP) algorithm for equalization to ensure low-latency processing by GPU and real-time throughput performance analysis. We emulate doubly-selective channels to analyse the performance of OTFS in a controlled environment. We compare the performance of OTFS and OFDM with various levels of time and frequency selectivity. By experimental results, we demonstrate that OTFS is more resilient against the disruptive effects of doubly-selective channels than OFDM.    
\begin{figure*}[t!]
	\centerline{\includegraphics[width=7in,height=5in,keepaspectratio]{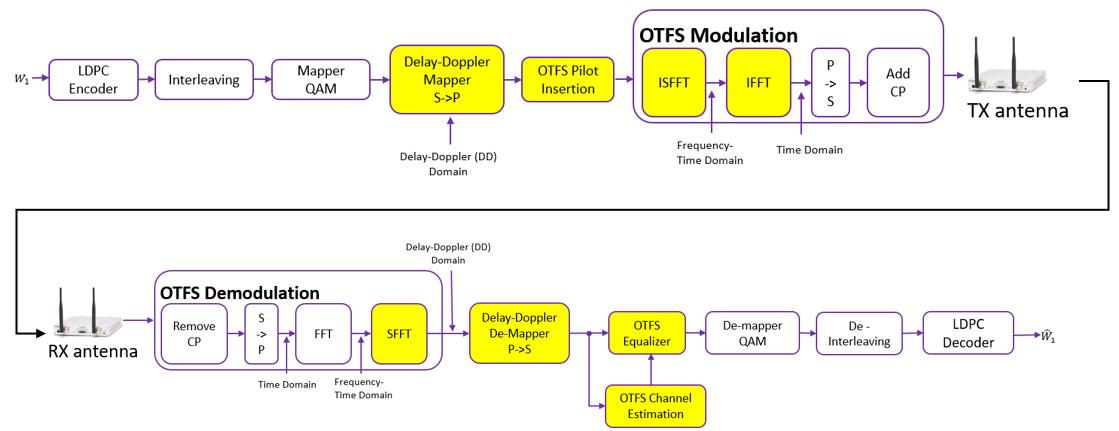}}
	\caption{OTFS transceiver architecture.}
	\label{fig:transceiver}
\end{figure*}

\section{System Model}
\label{sec:otfs}
In this section, we present the OTFS modulation expressions and the transceiver architecture. Next, we describe the resource allocation, air interface, and the channel model we consider in our experiments.   
\subsection{OTFS Modulation}
\label{eqn:modulation}
We describe the OTFS modulation technique in a Single-Input Single-Output (SISO) system \cite{hadani_2017, hadani_2017_2, shen_2019}. Let us denote the matrix $\mathbf{X}^{\mathrm{DD}} \in \mathbb{C}^{N_{Del} \times N_{Dopp}}$, which is the 2D data matrix used to place data and pilot symbols in delay-Doppler domain. The dimensions $N_{Del}$ and $N_{Dopp}$ are the numbers of resource units along the delay dimension and Doppler dimension, respectively. The data matrix is first applied an inverse symplectic finite Fourier transform (ISFFT), which can be expressed as 
\begin{align}
    \mathbf{X}^{\mathrm{ISFFT}}=\mathbf{F}_{N_{Del}}\mathbf{X}^{\mathrm{DD}}\mathbf{F}^{H}_{N_{Dopp}},
\end{align}
where $\mathbf{F}_{N_{Del}} \in \mathbb{C}^{N_{Del} \times N_{Del}}$ and $\mathbf{F}_{N_{Dopp}} \in \mathbb{C}^{N_{Dopp} \times N_{Dopp}}$ are fast Fourier
transform (FFT) matrices. A windowing is applied on the resulting signal in frequency-time domain, which can be written as 
\begin{align}
    \mathbf{X}^{\mathrm{FT}}=\mathbf{X}^{\mathrm{ISFFT}} \odot \mathbf{T}.
\end{align}
The resulting matrix is then passed through an OFDM demodulator and the 2D transmit signal block is obtained as 
\begin{align}
    \mathbf{S}=\mathbf{F}^{H}_{N_{Del}}\mathbf{X}^{\mathrm{FT}},
\end{align}
where $\mathbf{S}=[\mathbf{s}_{1}, \mathbf{s}_{2}, \ldots, \mathbf{s}_{N_{Dopp}}]$ and $\mathbf{s}_{i} \in \mathbb{C}^{N_{Del} \times 1}$ is an OFDM symbol $\forall i \in \{1, 2, \ldots, N_{Dopp}\}$.

When an OFDM modulator is employed to process the signal $\mathbf{X}^{\mathrm{FT}}$, it adds a Cyclic Prefix (CP) between the OFDM symbols $\mathbf{s}_{i}$ to avoid inter-symbol interfence (ISI) \cite{shen_2019}. However, a single CP at the beginning of the signal is sufficient to avoid ISI in OTFS, which makes it more efficient than OFDM in terms of overhead \cite{gaudio_2020, wei_2021}. Accordingly, the transmit signal $\mathbf{s} \in \mathbb{C}^{(N_{Del}N_{Dopp}+N_{CP}) \times 1}$ is obtained as 
\begin{align}
     \mathbf{s}=\mathbf{A}_{CP}\mathrm{vec}(\mathbf{S}).
\end{align}
The term $N_{CP}$ is the length of CP in samples and $A_{CP}=[\mathbf{D}_{CP}^{T}, \  \mathbf{I}_{N_{Del}N_{Dopp}}]^{T}$ is the CP addition matrix, where $\mathbf{I}_{N_{Del}N_{Dopp}}$ is the identity matrix of dimension $N_{Del}N_{Dopp} \times N_{Del}N_{Dopp}$ and $\mathbf{D}_{CP}$ is formed by taking the last $N_{CP}$ rows of $\mathbf{I}_{N_{Del}N_{Dopp}}$. The function $\mathrm{vec}(\mathbf{S})$ forms a column vector from the columns of matrix $\mathbf{S}$.

\subsection{Transceiver Architecture}
\label{sec:transceiver}
In this section, we present the implemented transceiver architecture employing OTFS modulation in a SISO system. 
Figure \ref{fig:transceiver} shows the block diagram of the implemented architecture.
At the transmitter, the input data bits $\mathbf{w}_{1}$ are first encoded using a Low-Density Parity-Check (LDPC) encoder for Forward Error Correction (FEC). The encoded bits are then interleaved and mapped onto Quadrature Amplitude Modulation (QAM) symbols. The QAM symbols are allocated to a resource grid in the delay-Doppler domain using a resource grid mapper, followed by the insertion of pilot symbols for channel estimation and synchronization. The OTFS process involves converting the signal to the frequency-time domain using the Inverse Symplectic Finite Fourier Transform (ISFFT), converting the resulting signal to the time domain using the Inverse Fast Fourier Transform (IFFT), transforming the parallel data stream into a serial data stream through parallel-to-serial conversion, and appending a cyclic prefix to mitigate inter-symbol interference (ISI) caused by multipath propagation. Finally, the modulated signal is sent to the USRP for transmission over the air.

Upon receiving the transmitted signal, the receiver begins with the OTFS demodulation process, which involves removing the cyclic prefix, converting the serial data stream back into a parallel data stream through serial-to-parallel conversion, performing a Fast Fourier Transform (FFT) to convert the signal to the frequency-time domain, and applying a Symplectic Finite Fourier Transform (SFFT) to convert the resulting signal to the delay-Doppler domain. The signal then goes through a delay-Doppler demapper, followed by an OTFS channel estimation and equalization module.
The equalized signal is then demapped using a QAM demapper, de-interleaved, and decoded with an LDPC decoder to recover the received bits.
\subsection{Channel Estimation and Equalization}
Each OTFS packet is transmitted with a dedicated pilot to help the receiver estimate the channel. In OTFS, all symbols are spread in time and frequency, and thus, all symbols experience the same channel in delay-Doppler domain \cite{hadani_2017_2}. Channel estimation is performed over an embedded impulse pilot in the delay-Doppler domain. In order to avoid the interference from adjacent data symbols, a zero region is applied, where the length of this region can be set according to the channel Doppler and delay spread \cite{Raviteja_2019}. 
The output of the channel estimation is fed into the MP equalizer proposed in \cite{Raviteja_2018}, which is slightly modified to consider the location of the pilot signal. The MP equalizer is a good match to be implemented on the GPU as it can perform the highly parallelized calculations efficiently. 
\begin{figure}[t!]
	\begin{subfigure}{.5\textwidth}
		\centerline{\includegraphics[width=3.2in,height=3.2in,keepaspectratio]{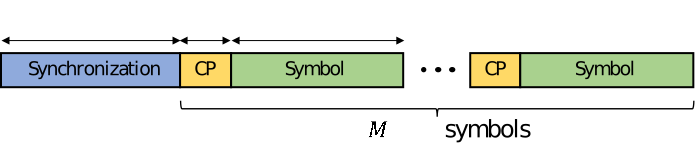}}
		\caption{OFDM.}
		\label{fig:slot_ofdm}
            \vspace{0.2cm}
	\end{subfigure}
        \newline
	\begin{subfigure}{.5\textwidth}
		\centerline{\includegraphics[width=3.2in,height=3.2in,keepaspectratio]{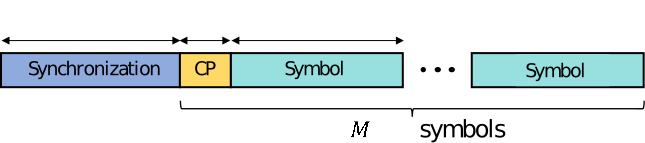}}
		\caption{OTFS.}
		\label{fig:slot_otfs}
	\end{subfigure}
	\caption{Time slot structure for transmission with OFDM and OTFS.}
        \label{fig:timeslots}
\end{figure}

\subsection{Resource Allocation and Air Interface}
\label{sec:resourceallocation}
In this section, we describe the air interface used for transmission with OFDM and OTFS. Fig.~\ref{fig:timeslots} depicts the time slot structures for both types of transmissions. A single-carrier synchronization signal of duration $T_{SYNC}$ seconds is appended at the start of each slot. The symbols of the synchronization signal are pulse-shaped with Root Raised Cosine (RRC) filter. For transmission with OFDM, a CP is appended before each symbol with duration $T_{CP}$ and $T_{OFDM}$, respectively. For transmission with OTFS, a single CP appended before symbol with duration $T_{CP}$ and $T_{OFDM}$, respectively. 
\begin{figure}[t!]
\centerline{\includegraphics[width=3.5in,height=4in,keepaspectratio]{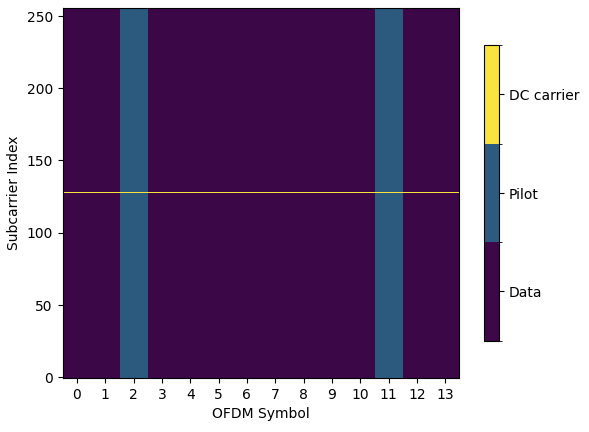}}
	\caption{OFDM resource grid.}
	\label{fig:ofdm_grid}
\end{figure}

\begin{figure}[t!]
	\centerline{\includegraphics[width=5in,height=3in,keepaspectratio]{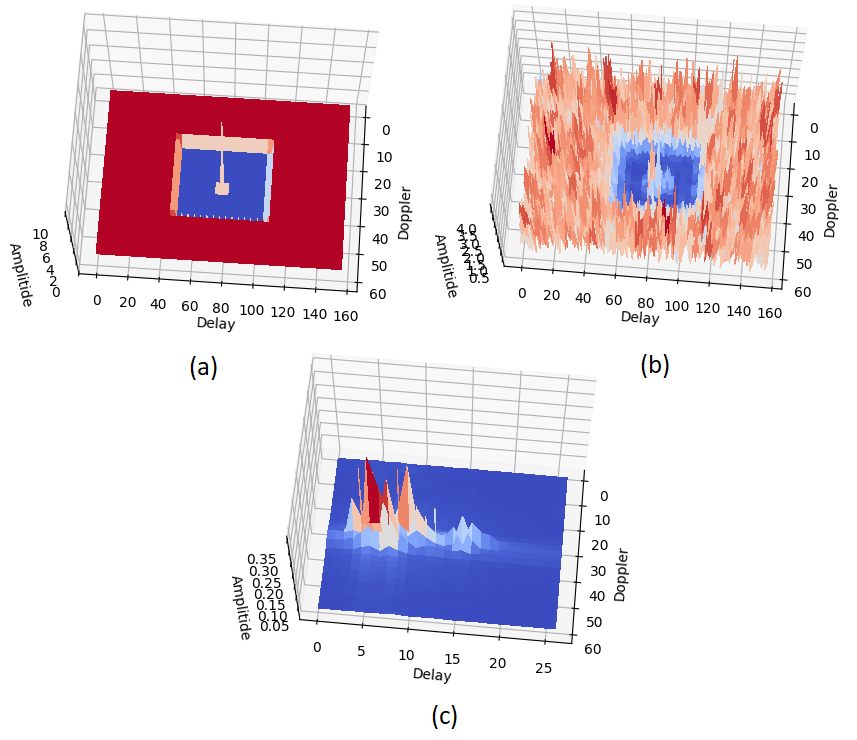}}
	\caption{(a) QPSK data in the OTFS resource grid. (b) The grid at the receiver and after applying the channel. (c) The channel response in DD domain}
	\label{fig:otfs_grid}
\end{figure}

\begin{figure*}[t!]
	\centerline{\includegraphics[width=7in,height=3in,keepaspectratio]{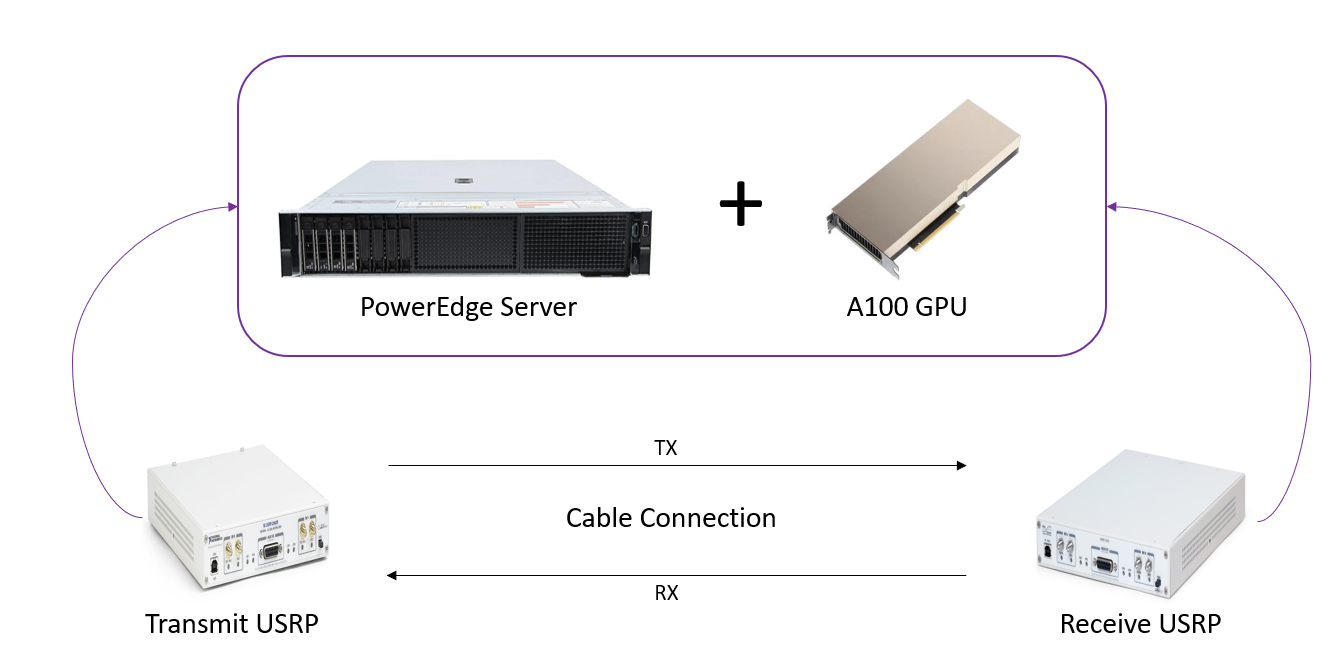}}
	\caption{Experimental setup.}
	\label{fig:setup}
\end{figure*}

For transmissions with OFDM, a total of $M_{OFDM}$ symbols are transmitted in a single time slot consisting of $N_{OFDM}$ subcarriers each. Pilots are embedded in $M_{pilot}$ symbols over all subcarriers. An example resource grid used for OFDM transmission is given in Fig.~\ref{fig:ofdm_grid} for $M_{OFDM}=14$, $N_{OFDM}=256$, and $M_{pilot}=2$. In the considered example, OFDM symbols $2$ and $11$ are used for pilot transmission while the rest of the symbols carry data symbols in all subcarriers except the DC subcarrier.

For transmissions with OTFS, a total of $M_{OTFS}=N_{Dopp}$ symbols are transmitted, obtained from a resource grid of $N_{Del}$ and $N_{Dopp}$ bins in delay-Doppler domain. A grid of $N_{Del,p}$ and $N_{Dopp,p}$ bins is left blank for pilot transmission, while the rest of the bins are used for data symbols. 
An example grid for OTFS transmission is given in Fig.~\ref{fig:otfs_grid} for $N_{Del}=60$, $N_{Dopp}=156$, $N_{Del,p}=58$, and $N_{Dopp,p}=32$.

\begin{figure}[t!]
	\begin{subfigure}{.5\textwidth}
		\centerline{\includegraphics[width=3.2in,height=3.2in,keepaspectratio]{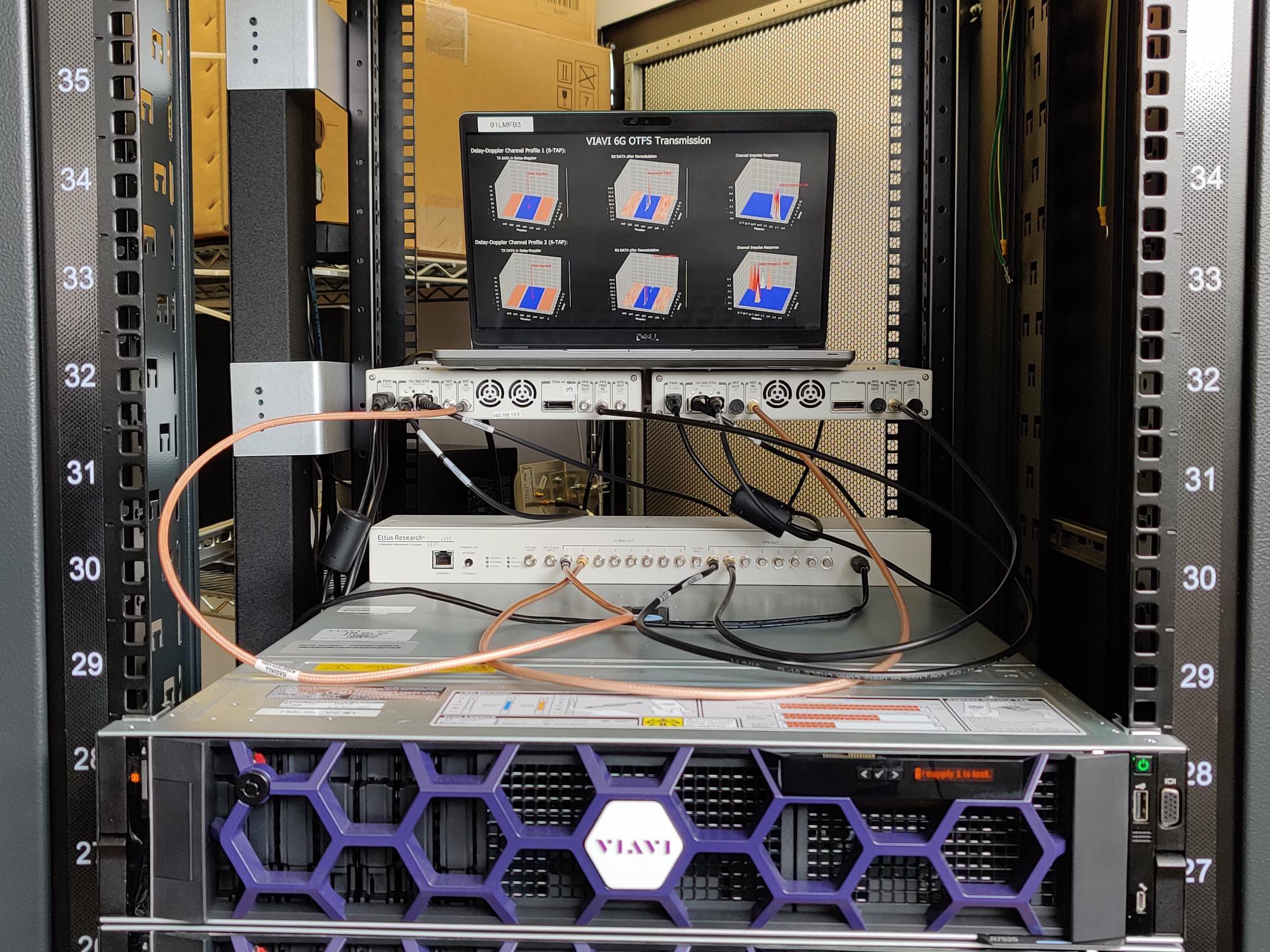}}
		\caption{Front view.}
		\label{fig:frontview}
            \vspace{0.2cm}
	\end{subfigure}
        \newline
	\begin{subfigure}{.5\textwidth}
		\centerline{\includegraphics[width=3.2in,height=3.2in,keepaspectratio]{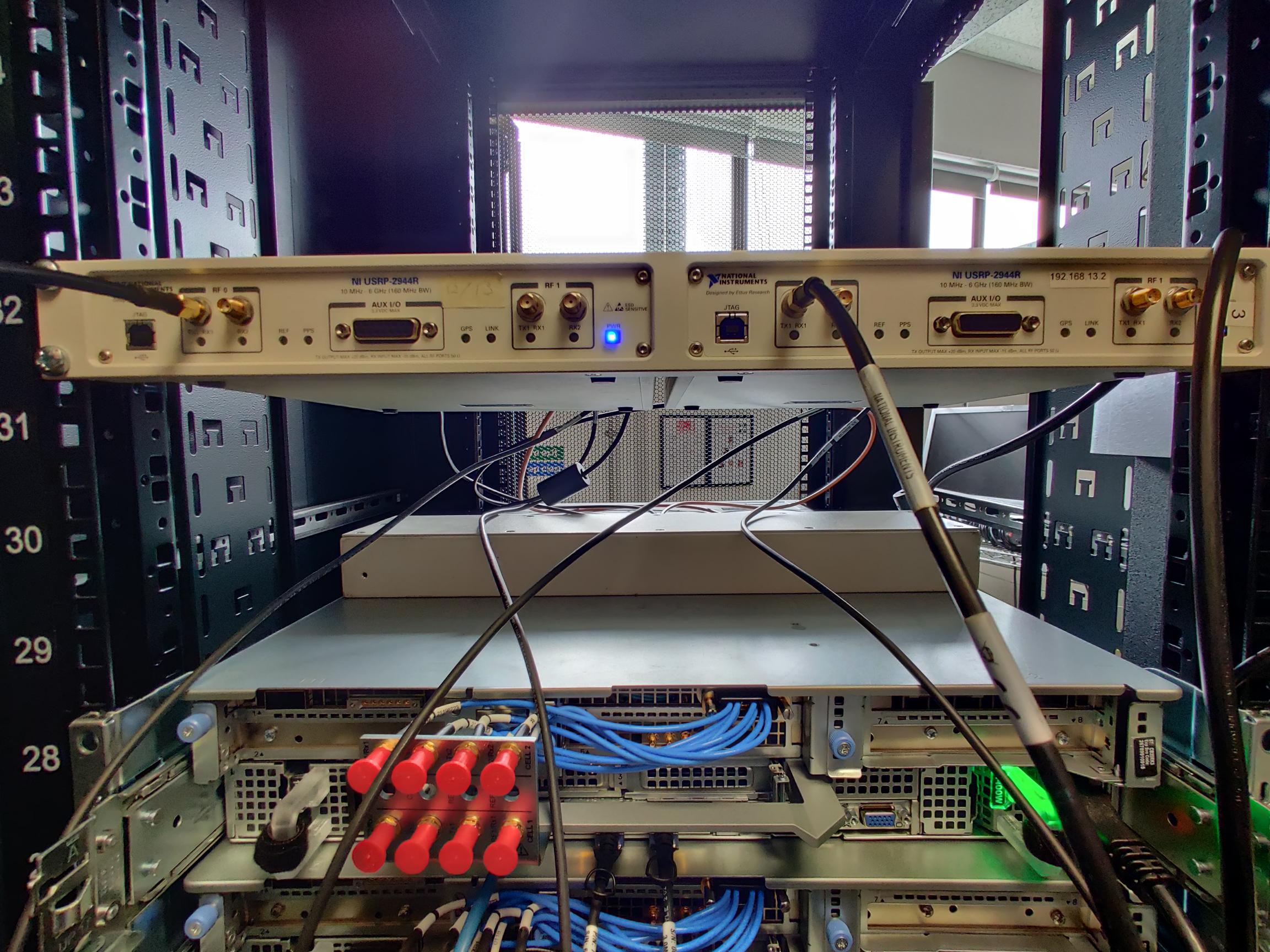}}
		\caption{Back view.}
		\label{fig:backview}
	\end{subfigure}
	\caption{The testbed, consisting of two USRP X310 devices, OctoClock CDA-2990, Dell PowerEdge R7525 server, and NVIDIA A100 PCIe Gen4 GPU.}
        \label{fig:front_back}
\end{figure}

\subsection{Channel Model}
\label{sec:channel}
We consider doubly selective channels to model a wireless medium with multiple paths and Doppler \cite{ma_2003}. The considered baseband channel model can be expressed in the time-delay domain as \cite{hlawatsch_matz_2011}:
\vspace{-0.3cm}
\begin{IEEEeqnarray}{rCl}
  c(t,\tau)  &=& \sum_{l=1}^L \alpha_l e^{j2\pi\nu_l t}  \delta(\tau-\tau_l), \quad l \in \left \lbrace 1,2,\ldots, L \right \rbrace,
\label{eq:channelModel}   
\vspace{-0.2cm}
\end{IEEEeqnarray}
where $\alpha_l$ is the complex attenuation factor, $\tau_l$ is the time delay, and $\nu_l$ is the Doppler frequency shift associated with the $l-\text{th}$ discrete propagation path. 

\section{Experimental Setup}
\label{sec:solution_perfectcsit}
In this section, we describe our testbed setup used to evaluate the OTFS transceiver architecture described in Section~\ref{sec:transceiver}. Fig.~\ref{fig:setup} shows the setup with a transmitter-receiver pair performing a point-to-point transmission. 
\subsection{Hardware Components}
In our experimental setup, we employ two Universal Software Radio Peripheral (USRP) X310 devices, one designated as the transmitter and the other as the receiver. 
To ensure accurate synchronization between the transmitter and receiver, we incorporate an OctoClock CDA-2990. Both USRP X310 devices are connected to the OctoClock via SMA cables for clock synchronization, ensuring precise alignment of the transmitted and received signals.
        
The setup is powered by a Dell PowerEdge R7525 server, equipped with two AMD EPYC 7763 64-Core processors, providing a total of 128 cores and 256 threads. These processors have a base frequency of 2.45 GHz. The system has a memory capacity of 256GB RAM and runs on Ubuntu 20.04 LTS. The USRP X310 devices are connected to the server through SFP+ ports using SFP+ cables, enabling high-speed data transfer between the TX and RX USRPs and the server, which helps to reduce the computing latency. 
The signal processing at both the transmitter and receiver sides is performed by an NVIDIA A100 PCIe Gen4 Graphical Processing Unit (GPU) with 80GB memory. A100 features 6,912 CUDA cores, 432 3rd generation Tensor Cores, a memory bandwidth of 2 TB/s, and a peak bandwidth of 64 GB/s, enabling faster communication between the GPU and the host system, which is essential for handling complex signal processing tasks in real-time. 
%It delivers 19.5 TFLOPs of single-precision performance, 624 TOPs (INT8) and 312 TFLOPs (FP16) of AI performance, and 39.5 TFLOPs of ray tracing performance. 
%The A100 PCIe Gen4 GPU provides a peak bandwidth of 64 GB/s, enabling faster communication between the GPU and the host system, which is essential for handling complex signal processing tasks in real-time. 
%This makes it an excellent choice for applications in fields like machine learning, artificial intelligence, high-performance computing, and data analytics. 
To further enhance parallel processing capabilities, Multiple Instance GPU (MIG) technology is employed on both the transmitter and receiver sides, enabling the application to run in parallel across multiple GPU instances and increase the GPU utilization. The setup and the connections are presented in Fig.~\ref{fig:front_back}.

The experiments are conducted using a cable-based connection between the transmitter and receiver USRPs. This approach allows us to eliminate the uncertainties and potential interference associated with external wireless communication, ensuring a controlled environment for the experiments. The doubly-selective channels are emulated on GPU by means of the software environment, which is detailed in the next section.

%\begin{figure*}[htbp]
%	\centerline{\includegraphics[width=7in,height=5in,keepaspectratio]{OTFS_3D.PNG}}
%	\caption{OTFS transmission.}
%	\label{fig:otfs_channel}
%\end{figure*}

\subsection{Implementation Software Environment}
In this section, we provide an overview of the software environment used to implement the transceiver modules on GPU and establish the communication between the GPU and the USRPs. 
We use Sionna to implement the blocks in our transceiver architecture \cite{hoydis_2022}. Sionna is a TensorFlow based software library designed for link-level simulation of communication systems. 
%It offers a wide range of pre-built modules and functions that streamline the design, simulation, and implementation of complex communication systems, enabling researchers and engineers to focus on the innovation and optimization of their designs.

%One of the primary reasons for choosing the Sionna library for our project is its versatility and adaptability to various communication scenarios. The library is designed with a modular approach, making it easy to mix and match different components based on the specific requirements of a given system. Furthermore, Sionna provides a high level of abstraction, which allows for seamless integration with other software tools and hardware platforms, such as the USRP devices used in our experimental setup.

In our implementation, we employ several modules from the Sionna library. At the transmitter side, we use the modules RRC filter for pulse shaping of synchronization signal, FFT and IFFT for time-frequency domain conversions, CP module for mitigating inter-symbol interference, QAM module for symbol mapping LDPC module for FEC.
%, which have facilitated the efficient realization of our transceiver architecture. 
%The first block in Figure 5 is the PSS (Primary Synchronization Signal) synchronizer, responsible for ensuring accurate timing synchronization between the transmitter and receiver. 
%This is a crucial component for maintaining the integrity of the transmitted signal and minimizing potential errors due to misalignment.

%Following the PSS synchronizer at the transmitter is the OTFS (Orthogonal Time Frequency Space) signal block, which serves as the core of our transceiver architecture. Within the OTFS block, we employ several Sionna library modules, such as RRC filter for pulse shaping of synchronization signal, FFT and IFFT for time-frequency domain conversions, CP module for mitigating inter-symbol interference, QAM module for symbol mapping LDPC module for FEC.

At the receiver end, a correlator is implemented to perform correlation with the received synchronization signal to accurately synchronize and detect the transmitted information. Following the successful correlation and synchronization, the receiver proceeds to process the received signal using Sionna modules to perform channel estimation, equalization, and decoding to recover the original transmitted data. 
%The seamless integration of these components in both the transmitter and receiver ensures efficient processing and reliable communication.

%In the Sionna modules, we leverage TensorFlow as the underlying framework for implementing signal processing algorithms. 
%TensorFlow provides a high-level, flexible interface for defining and executing computational graphs, which can be optimized and parallelized on GPUs.

%By leveraging the Sionna library and its wide array of modules, we have been able to effectively implement our transceiver architecture, which is designed to exploit the benefits of the Orthogonal Time Frequency Space (OTFS) modulation technique in Single-Input Single-Output (SISO) communication systems. 

%It is important to note that while a significant portion of the modules used in our implementation are based on the Sionna library, some custom modules and adaptations have also been employed to address the unique requirements of our experimental setup. 

In order to implement new modules for OTFS modulation and demodulation, we use TensorFlow, which enables parallel computation and seamless integration of the custom modules into the Sionna library.
%we follow a systematic approach that involves identifying the specific requirements, designing the module architecture, and coding the module using TensorFlow operations and functions. 
%This approach enables efficient computation and seamless integration of custom modules into the Sionna library. 
After implementing the custom modules, we employ graph execution to further enhance the performance and adaptability of our system. This technique allows for the efficient execution of computational graphs and optimizes resource utilization, ultimately leading to a more robust and versatile transceiver architecture.
To communicate with the USRP devices, we use the Universal Hardware Driver (UHD) library. %This library allows integration and control of USRP devices in our experimental setup. 
\begin{table}[t!]
\vspace{0.3cm}
\caption{Parameters}
\centering
 \begin{tabular}{|c |c | c| c|} 
 \hline
 Parameter & Value & Parameter & Value \\ [0.5ex] 
 \hline\hline
 $T_{SYNC}$ ($\mu$s) & 8.33 & $N_{OFDM}$ & $256$ \\ 
 \hline
 $T_{CP}$ ($\mu$s) & 1.56 & $M_{pilot}$ & 2 \\
 \hline
 $T_{OFDM}$ ($\mu$s) & 33.3/66.6 & $N_{Dopp}$ & 15 \\
 \hline
 $T_{OTFS}$ (ms)& 1 & $N_{Del}$ & 330 \\
 \hline
 $M_{OFDM}$ & $14$ & $N_{Dopp,p}$ & 15 \\
 \hline
 $M_{OTFS}$ & 15 & $N_{Del,p}$ & 26 \\ [1ex] 
 \hline
 \end{tabular}
 \label{table:parameters}
\end{table}

\section{Experimental Results}
\label{sec:numerical}
In this section, we present the experimental evaluation of OTFS modulation in our setup. The parameters used in our experiments are given in Table~\ref{table:parameters}.

The synchronization signal is obtained by applying RRC pulse-shaping to the Primary Synchronization Signal (PSS) in 5G New Radio (NR) \cite{3gpp}. The PSS signal consists of $127$ symbols and the applied RRC filter has a roll-off factor of $0.5$. 
\begin{figure}[t!]
    \vspace{0.3cm}
	\centerline{\includegraphics[width=3.2in,height=3.2in,keepaspectratio]{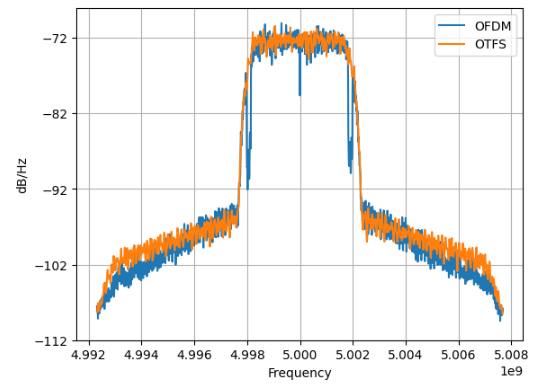}}
	\caption{Power spectral density of OFDM and OTFS.}
	\label{fig:psd}
\end{figure}

We start by comparing the generated OTFS and OFDM signals. Fig.~\ref{fig:psd} shows the comparison of Power Spectral Densities (PSDs) of OFDM and OTFS modulated signals at a carrier frequency of $5$~GHz. The plot showcases the frequency domain characteristics under the same channel conditions, with a bandwidth that is consistent across both PSDs. 

Next, we compare the throughput performance of OTFS and OFDM. We fix the modulation-coding scheme (MCS) to QPSK and a coding rate of $1/3$ for both transmissions. The experiments are conducted at a Signal-to-Noise Ratio (SNR) of $30$dB to minimize the affect of noise on the resulting performance and observe the variations in performance due to doubly-selective channel effects only. We use the Tapped Delay Line (TDL) channel model \cite{3gpp_2} emulated in Sionna with $23$ taps, a delay spread of $100$nsec and various user speed to analyse the performance.

\begin{figure}[t!]
	\centerline{\includegraphics[width=3.2in,height=3.2in,keepaspectratio]{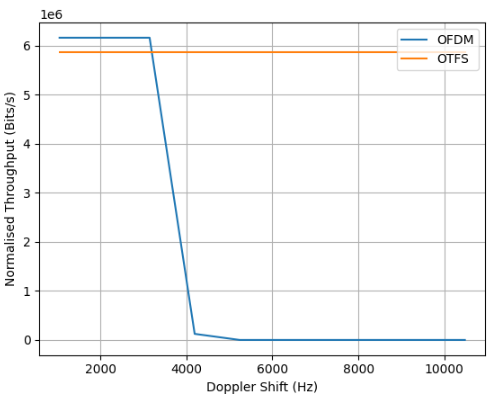}}
	\caption{Subcarrier spacing of 15KHz.}
	\label{fig:scs_15}
\end{figure}

\begin{figure}[t!]
	\centerline{\includegraphics[width=3.2in,height=3.2in,keepaspectratio]{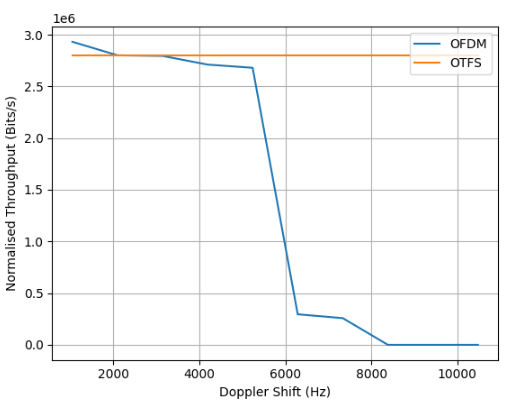}}
	\caption{Subcarrier spacing of 30KHz.}
	\label{fig:scs_30}
\end{figure}

Fig.~\ref{fig:scs_15} and \ref{fig:scs_30} demonstrate the normalized throughput obtained using OTFS and OFDM for subcarrier spacing of $15$kHz and $30$kHz with respect to Doppler shift. One can see that the throughput of OFDM stays almost constant in both scenarios due to operating at high SNR up to a certain Doppler value. After a certain Doppler frequency is reached, the throughput of OFDM drops to zero as the inter-carrier interference (ICI) becomes high enough to reduce the achievable rate below what the MCS scheme can support. On the other hand, OTFS achieves a robust performance against Doppler shift owing to its delay-Doppler domain processing. 

It is essential to underscore that the portion of the resource allocated to pilots and the signal, as well as the manner in which the pilot signal is embedded into the data grid, are critical factors that can influence throughput. Furthermore, the overhead introduced by the cyclic prefix (CP) can notably influence the starting point of the throughput for each scheme.

\section{Conclusion}
\label{sec:conclusion}
In this work, we demonstrate a successful implementation and evaluation of OTFS modulation in a real-time testbed consisting of a GPU and USRPs. By leveraging the Sionna library and using TensorFlow, we implemented a low-latency transceiver architecture, which is designed to take full advantage of the OTFS scheme. By means of our experimental setup, we highlighted the potential of OTFS to provide significant performance improvements in challenging communication environments, such as those with high mobility and severe multipath conditions. 

For future work, there are several possible directions to explore. Firstly, the performance of the OTFS could be further investigated under different channel conditions and scenarios, such as varying levels of interference and noise, and transitioning from cable-based to over-the-air experiments. Additionally, different OTFS algorithms and variants can be analyzed and compared to identify the most suitable options for specific communication scenarios. Moreover, the design and performance of OTFS can be investigated in a broader context of multiple access technology and multiuser communication systems. This would provide valuable insights into the applicability of OTFS in various real-world scenarios and enable the development of more efficient and robust communication strategies.
%Lastly, the design and performance of OTFS in a multi-input multi-output (MIMO) systems can be investigated. 
%\textcolor{red}{Secondly, the implementation of advanced error correction and detection mechanisms, in addition to the currently employed LDPC coding, could enhance the overall robustness and reliability of the communication system. (WHAT DO WE MEAN HERE?)}
%By addressing these potential research areas, we aim to continue enhancing the performance of OTFS-based communication systems and contribute to the development of future wireless communication technologies.

%\begin{itemize}
	%\item Conclusion and future work
%\end{itemize}	

\end{document}